\documentclass[aps,pra,twocolumn,showpacs,amsmath,amssymb,floatfix,footinbib,superscriptaddress]{revtex4-1}
\usepackage{graphicx}
\usepackage{amsmath}
\usepackage{latexsym}
\usepackage{amsmath}
\usepackage{amssymb}
\usepackage{float}
\usepackage{epstopdf}
\usepackage{bm}
\usepackage{amsfonts}
\usepackage[ansinew]{inputenc}
\usepackage[usenames,dvipsnames]{pstricks}
\usepackage{subfigure}
\usepackage{epsfig}
\usepackage{longtable}
\usepackage{url}
\usepackage{hyperref}
\usepackage{color}
\definecolor{darkblue}{rgb}{0.0,0.0,0.3}
\hypersetup{colorlinks,breaklinks,linkcolor=blue,urlcolor=blue,anchorcolor=blue,citecolor=blue}

\begin{document}
\baselineskip=14pt
\title{Study of Rydberg blockade mediated optical non-linearity in thermal vapor using optical heterodyne detection technique}
\author{Arup Bhowmick}
\email{E-mail: arup.b@niser.ac.in}
\affiliation{School of Physical Sciences, National Institute of Science Education \& Research,
Bhubaneswar 751005, INDIA.}
\author{Dushmanta Kara}
\affiliation{School of Physical Sciences, National Institute of Science Education \& Research,
Bhubaneswar 751005, INDIA.}
\author{Ashok K. Mohapatra}
\affiliation{School of Physical Sciences, National Institute of Science Education \& Research,
Bhubaneswar 751005, INDIA.}

%\maketitle
\date{\today}
\begin{abstract}
We demonstrate the phenomenon of blockade in two-photon excitations to the Rydberg state in thermal vapor. A technique based on optical heterodyne is used to measure the dispersion of a probe beam far off resonant to the D$2$ line of rubidium in the presence of a strong laser beam that couples to the Rydberg state via two-photon resonance. Density dependent suppression of the dispersion peak is observed while coupling to the Rydberg state with principal quantum number, $n=60$. The experimental observation is explained using the phenomenon of Rydberg blockade. The blockade radius is measured to be about $2.2$ $\mu$m which is consistent with the scaling due to the Doppler width of 2-photon resonance in thermal vapor. Our result promises the realization of single photon source and strong single photon non-linearity based on Rydberg blockade in thermal vapor.
\end{abstract}
\maketitle

\section{introduction}
Rydberg atoms are enriched with enhanced two-body interactions. When atoms in a dense frozen ensemble are excited to the Rydberg state using narrow band lasers, strong Rydberg-Rydberg interactions lead to excitation blockade. This blockade interaction generates a highly entangled many-body quantum state in an ensemble of atoms which has become the basis for fundamental quantum gates using atoms~\cite{jaks00,luki01,isen10,wilk10} or photons~\cite{frie05} and for realization of single photon source~\cite{saff02,dudi12}. Rydberg blockade in ultra-cold atoms and Bose-Einstein condensate (BEC) has been proposed to study strongly correlated systems~\cite{weim08,pohl09,pupi10,henk10}. In addition, strong photon-photon interactions enabled by Rydberg blockade mediated optical non-linearity has been proposed~\cite{sevi11,gors11}. Rydberg blockade has been demonstrated for an ensemble of ultra-cold atoms in a magneto-optical trap (MOT)~\cite{tong04,sing04,cube05,vogt06} or in a magnetic trap~\cite{heid07,rait08} and also for single atoms trapped in optical micro traps~\cite{urba09,gaet09}. Recently, strong optical non-linearity mediated by Rydberg blockade in cold atom has been observed for weak classical light~\cite{prit10} as well as for single photons~\cite{peyr12,firs13} and also in a cold atomic sample inside an optical cavity~\cite{pari12}.

\begin{figure}[t]
\begin{center}
\epsfig{file=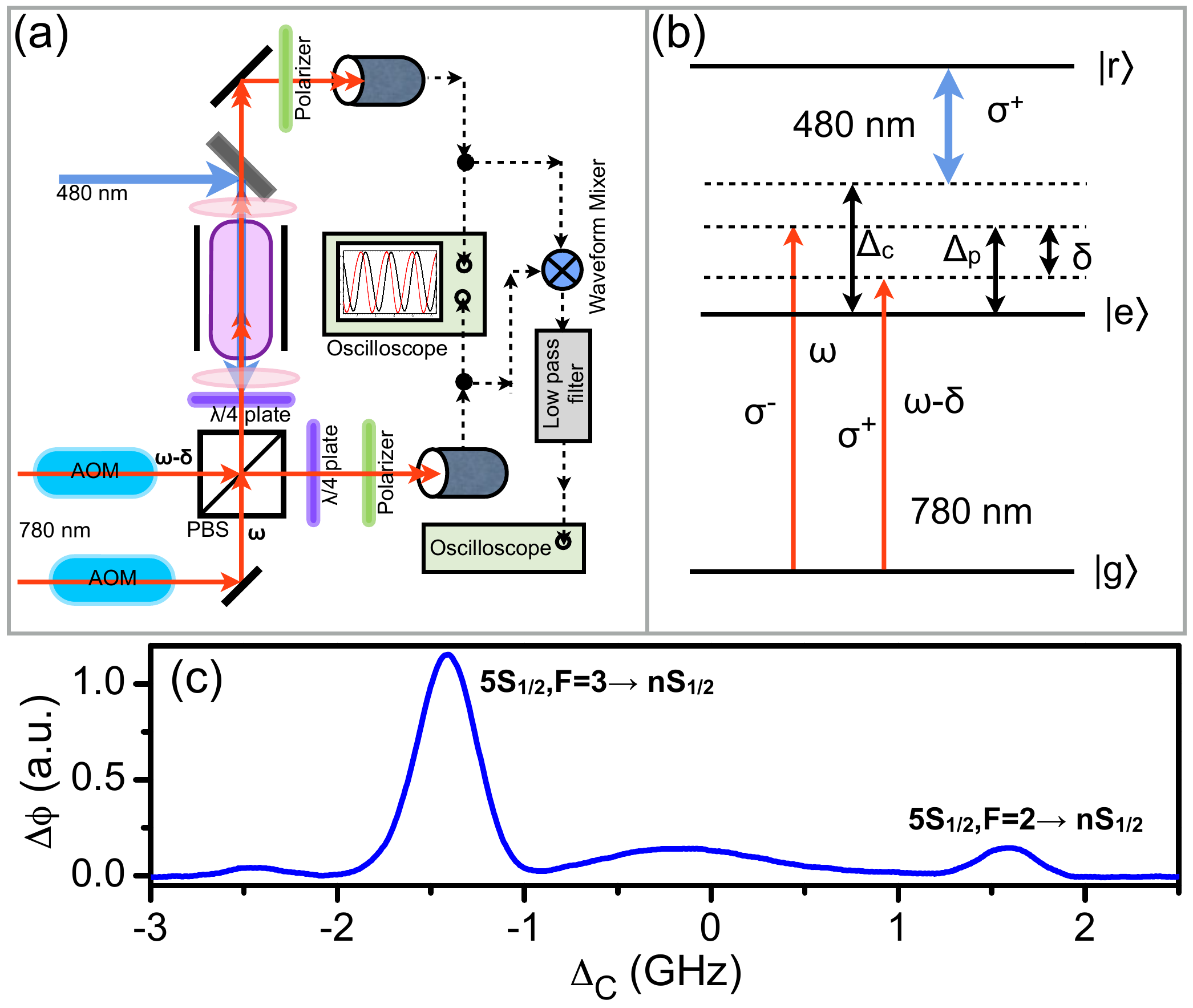,clip=,angle=0,width=8.0cm}
\caption[]{(a) Schematic of the experimental set up. (b) Energy level diagram for 2-photon 
transition to the Rydberg state. Two probe beams with $\sigma^+$ and $\sigma^-$ polarizations couple the transition 
$5$s$_{1/2}$, F$=3$ $\left(\left|g\right\rangle\right)$ $\longrightarrow$ $5$p$_{3/2}$ $\left(\left|e\right\rangle\right)$ of $^{85}$Rb. 
The coupling laser with $\sigma^+$ polarization couples the transition $5$p$_{3/2}$ $\left(\left|e\right\rangle\right)$ $\longrightarrow$ 
$n$s$_{1/2}$ $\left(\left|r\right\rangle\right)$. The probe(coupling) detuning is $\Delta_p$$\left(\Delta_c\right)$ and frequency 
offset between the probe beams is $\delta$. (c) Typical dispersion spectrum of a probe beam by scanning the coupling over $5$ GHz.}
\end{center}
\label{fig1}
\end{figure}

The blockade radius for an ultra-cold atomic sample is defined as $r_b=\sqrt[6]{\frac{C_6}{\hbar\Omega_{eff}}}$, where $\Omega_{eff}$ 
is the effective Rabi frequency of the Rydberg excitation and $C_6$ is the strength of the van der Waals type Rydberg-Rydberg interaction.
For a thermal ensemble of atoms, the blockade radius is affected by the Doppler width ($\Delta\nu_D$) due to the thermal motion of 
atoms. Since the van der Waals interaction scales as the sixth power of the inter-atomic separation, the blockade 
radius is scaled down as $r_b\propto\sqrt[6]{\Delta\nu_D}$~\cite{kubl10}. For thermal rubidium atoms at room temperature, 
the blockade radius is decreased only by a factor in the range of $2$ $-$ $3$. If one works with a Rydberg state, $n>60$, 
the blockade radius in thermal vapor is of the order of a few microns. Electromagnetically induced transparency involving Rydberg state
(Rydberg EIT) has been demonstrated in thermal vapor cell~\cite{moha07} and in micron sized vapor cell~\cite{kubl10}. 
Van der Waals interactions of Rydberg atoms in thermal vapor has been observed recently~\cite{balu13}. In addition, four wave mixing 
involving Rydberg state~\cite{koll12} and large dc Kerr non-linearity of a Rydberg EIT medium has also been demonstrated in thermal 
vapor~\cite{moha08}.

In this article, we present the first ever strong evidence of Rydberg blockade in thermal vapor using narrow-band lasers for 
Rydberg excitations. The schematic of the experimental set up is shown in figure 1(a). A technique based on optical heterodyne 
is used to measure the dispersion of a probe beam due to two-photon excitation to the Rydberg state. A similar method used for 
dispersion measurement in atomic vapors is reported in references~\cite{akul99,akul04}. An external cavity diode laser operating 
at $780$ nm is used to derive two probe beams. A frequency offset is introduced between the probe beams by passing 
them through two acousto-optic modulators. Both the beams are made to superpose using a polarizing cube beam splitter (PBS). 
The interference beat signals are detected using two fast photo-detectors by introducing polarizers at both the output ports of 
the PBS. The probe beams coming out of one of the output ports of the PBS propagate through a magnetically shielded rubidium vapor 
cell. The density of $^{85}$Rb vapor can be increased to $5.0\times 10^{13}$ cm$^{-3}$ by heating the cell up to $130^0$C. 
The coupling beam is derived from a frequency doubled diode laser operating at $478 - 482$ nm and it counter-propagates the 
probe beams through the cell. The coupling and the probe beams are focused inside the cell using lenses. The waist and the 
Rayleigh range of the probe (coupling) beams are $35$ $\mu$m ($50$ $\mu$m) and $12$ mm ($10$ mm), respectively. 
The peak Rabi frequencies of the laser beams and their variations over the length of the vapor cell were calculated~\cite{rabi} 
using the same parameters of the beams and were included in the theoretical model to fit all the experimental data.          

The probe beams propagating through the medium can undergo different phase shifts by choosing suitable polarizations 
of the probe and the coupling beams. The polarization of the coupling beam is chosen to be $\sigma^{+}$ and the probe beams to be
$\sigma^{+}$ and $\sigma^{-}$ as shown in figure 1(b). The probe beam with $\sigma^{+}$ polarization can not couple the two-photon 
transition, $5$s$_{1/2}$ $\rightarrow$ $n$s$_{1/2}$ and doesn't go through any phase shift due to two-photon process. We use 
it as the reference probe beam. However, the other probe beam with $\sigma^{-}$ polarization can couple the same two-photon transition
and hence, goes through a phase shift due to two-photon excitation to the Rydberg state. This additional phase shift of the signal 
probe beam appears as a phase shift in the respective beat signal and is measured by comparing to the phase of the reference beat 
signal detected at the other output port of the PBS. Since, both the beat signals are the output of the same interferometer, the noise 
due to vibration and acoustic disturbances are strongly suppressed. The signal-to-noise ratio was further improved by using a lock-in 
amplifier. 

\begin{figure}[t]
\begin{center}
\epsfig{file=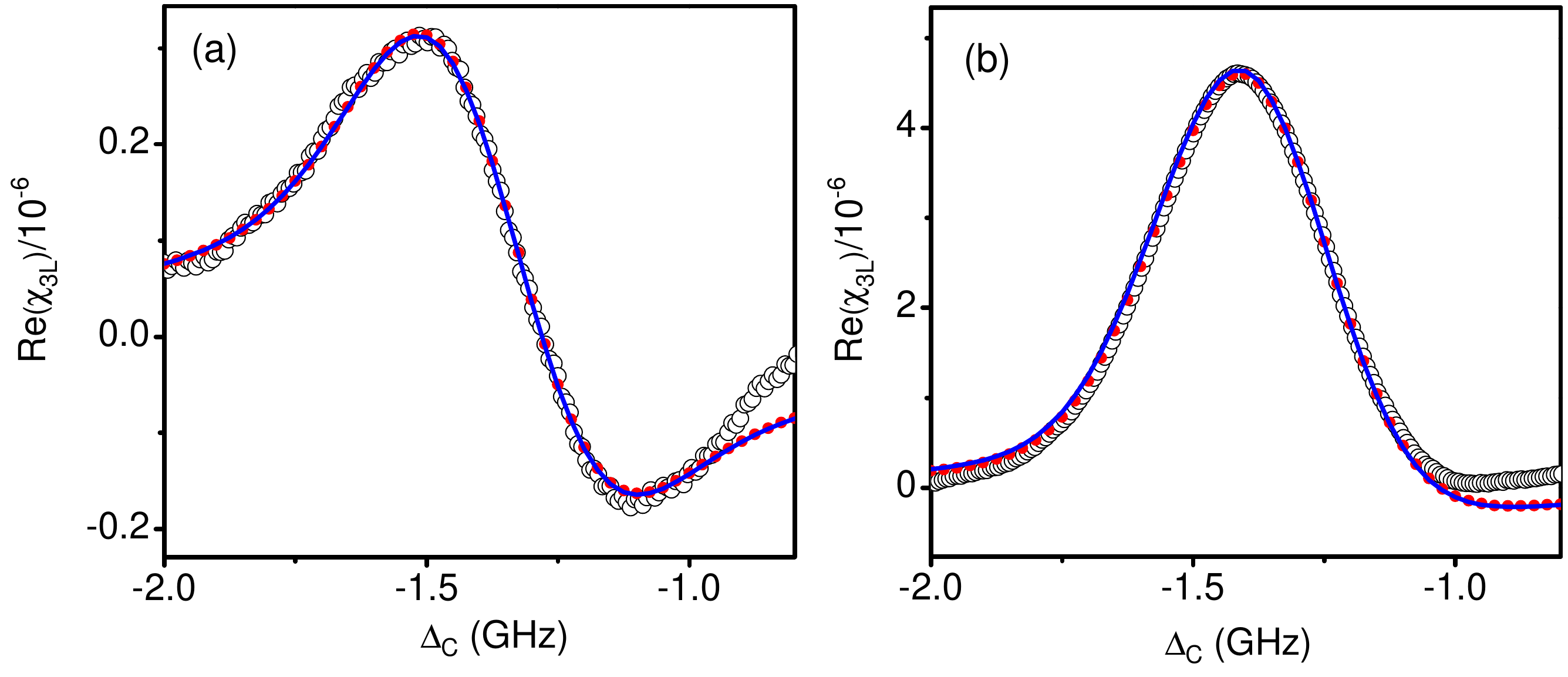,clip=,angle=0,width=8.5cm}
\caption[]{(a) Real part of the susceptibility of the medium due to 2-photon resonance. The peak Rabi frequencies of the coupling beam was $24$ MHz and of the probe beam was $60$ MHz (a) and $400$ MHz (b). The lasers are coupled to the Rydberg state ($n=30$s$_{1/2}$). Open circles ($\circ$) are the refractive index measured in the experiment, solid circles ($\bullet$) are the calculated refractive index using the exact model of a 3-level atom interacting with a probe and a coupling beams. Solid lines are the calculated refractive index using an approximate model of considering the 3-level atom as an effective 2-level atom. To compare with the theoretical model, the experimental data were scaled by a multiplication factor which can be accounted for overall gain in the experiment.}
\end{center}
\label{fig2}
\end{figure}

The beat signals passing through a high pass filter have the form, 
$D_{\alpha}=A_{\alpha}\cos(\delta t+\phi_{\alpha})$, $\alpha=r,s$. Here, $\delta$ is the frequency offset between the 
probe beams and $A_r$ ($A_s$), $\phi_r$ ($\phi_s$) are the amplitude and phase of the reference 
(signal) beat, respectively. $A_r$ and $A_s$ depend on the power of the probe beams falling on the detector.
The beat signals are then multiplied using an electronic waveform mixer and are passed through a 
low pass filter. The output of the low pass filter gives a DC signal, $S_{L}=2A_{r}A_{s}\cos(\Delta\phi)$ 
where $\Delta\phi=\phi_r-\phi_s$. For any small variation of the phase around $\Delta\phi=0$ ($\Delta\phi=\frac{\pi}{2}$), 
$S_L$ gives information about absorption (dispersion) of the probe beam propagating through the medium. A phase offset  
between the beams falling on the reference detector was introduced by placing a $\frac{\lambda}{4}$-plate before the 
polarizer. As a result, the phase of the reference beat can be controlled by rotating the polarizer axis. The refractive 
index of the $\sigma^-$ probe beam due to two-photon excitation to the Rydberg state 
can be measured with a phase offset, $\Delta\phi=\frac{\pi}{2}$. The detailed principle of the technique can be found in 
reference~\cite{bhow16}. For a larger frequency offset between the probe beams in comparison to their Rabi frequency, 
the experimental result matches well with the standard model of a 3-level atom interacting with a single probe and 
coupling beams~\cite{bhow16}. Hence, a large frequency offset of $800$ MHz is used in the experiment.

In the experiment, the probe beam was stabilized at $1.3$ GHz blue detuned to the D2 line of $^{85}$Rb. The coupling 
laser frequency was scanned to observe the dispersion of the probe beam by measuring its phase shift due to the two-photon 
excitations to the Rydberg state. A typical dispersion spectrum with the coupling beam scanning over $5$ GHz is shown in figure 1(c).
The 2-photon resonance peaks corresponding to the transitions, $5$s$_{1/2}$F$=3\longrightarrow$ $n$s$_{1/2}$ and 
$5$s$_{1/2}$F$=2\longrightarrow$ $n$s$_{1/2}$ are observed and are used to normalize the frequency axis. The dispersion peak 
corresponding to the $5$s$_{1/2}$F$=3\longrightarrow$ $n$s$_{1/2}$ transition was analyzed for the further study of Rydberg excitation. 
For a weak probe beam, an usual dispersion profile of the two-photon resonance is observed as shown in figure 2(a). 
However, an absorptive like dispersion profile is observed for a stronger
probe beam as shown in figure 2(b). In order to explain the shape of the dispersion profile, we consider a 
three level atomic system interacting with two monochromatic laser field in ladder configuration as shown in figure 1(b).
$\Omega_p$ and $\Omega_c$ are used as probe and coupling Rabi frequencies respectively.  
The density matrix equation, $i\hbar\dot{\hat{\rho}}=[\hat{H},\hat{\rho}]+i\hbar\mathcal{L}_D\hat{\rho}$
is solved numerically in steady state and is averaged over the Maxwell-Boltzmann velocity distribution of the atom. 
Here, $\hat{H}$ is the Hamiltonian for a three-level atom interacting with two mono-chromatic light field in a 
suitable rotating frame and $\mathcal{L}_D$ is the Lindblad operator which takes care of the decoherence in the system.
In the numerical calculation, we have used the decay rate of the channels $\left|e\right\rangle\rightarrow\left|g\right\rangle$ 
as $2\pi\times 6$ MHz. A population decay rate of the Rydberg state to the ground state denoted as $\Gamma_{rg}$ is used to account 
for the finite transit time of the thermal atoms in the laser beams. The dipole $\rho_{rg}$ dephases at a rate of 
$\frac{\Gamma_{rg}}{2}+\gamma_{rel}$, where $\gamma_{rel}$ accounts for the relative laser noise between the probe and the coupling beams. 
$\Gamma_{rg}$ and $\gamma_{rel}$ are of the order of $2\pi\times 1$ MHz. 
%The decay of Rydberg population due to its finite life time is less than $2\pi\times 100$ kHz and has negligible effect.% 
The dispersion of the probe beam due to 2-photon excitation calculated using the theoretical model is shown in figure 2. 

To get an insight to such unusual dispersion profile, we approximated the 3-level atom as an effective 2-level atom by 
adiabatically eliminating the intermediate state~\cite{han13}. Using the same approximation ($\Delta_p>>\Gamma_{eg},\Omega_p$ and 
$\rho_{ee}\approx 0$) in the steady state equations of 3-level atom, the susceptibility of the medium can be written as 
$\chi=\chi_{2L}+\chi_{3L}$, where $\chi_{2L}$ is the susceptibility of the lower transition without coupling beam and $\chi_{3L}$ 
is the susceptibility due to 2-photon resonance only. $Re\left(\chi_{2L}\right)=\frac{2N\left|\mu\right|^2}{\epsilon_0\hbar}
\left(\frac{-1}{\Delta_p}\right)$ where $N$ is the vapor density and $\mu$ is the dipole moment of the lower transition. In the regime $\Omega_p^2>>\Gamma_{eg}\Gamma_{rg}$,
\begin{equation}
Re\left(\chi_{3L}\right)=\frac{N\left|\mu\right|^2}{\epsilon_0\hbar}\left(\frac{1}{\Delta_p}\right)\left(1+\frac{2\Delta^2}{\Omega_p^2}-
\frac{4\Delta\Delta_p}{\Omega_p^2}\right)\rho_{rr}
\end{equation}
Where $\Delta=\Delta_2+\frac{\Omega_p^2}{4\Delta_s}-\frac{\Omega_c^2}{4\Delta_s}$ with $\Delta_2=\Delta_p+\Delta_c$ and 
$\Delta_s=\frac{\Delta_p-\Delta_c}{2}$. The Rydberg population ($\rho_{rr}$) can be determined analytically using the effective 
2-level model. Doppler averaging of the equation (1) with same laser parameters fits well with the experimental data as shown in 
figure 2. This approximate model shows a very little deviation from the exact 3-level calculation for the probe Rabi frequency 
up to $500$ MHz. It is worthwhile to mention that 
$Im\left(\chi_{3L}\right)=\frac{N\left|\mu\right|^2}{\epsilon_0\hbar}\left(\frac{2\Gamma_{rg}}
{\Omega_p^2}\right)\rho_{rr}$. Comparing it with $Re\left(\chi_{3L}\right)$, the dispersion peak is an order of magnitude larger 
than the absorption peak for $\Delta_p\approx 1$ GHz and $\Omega_p\approx 100$ MHz.

\begin{figure}[t]
\begin{center}
\epsfig{file=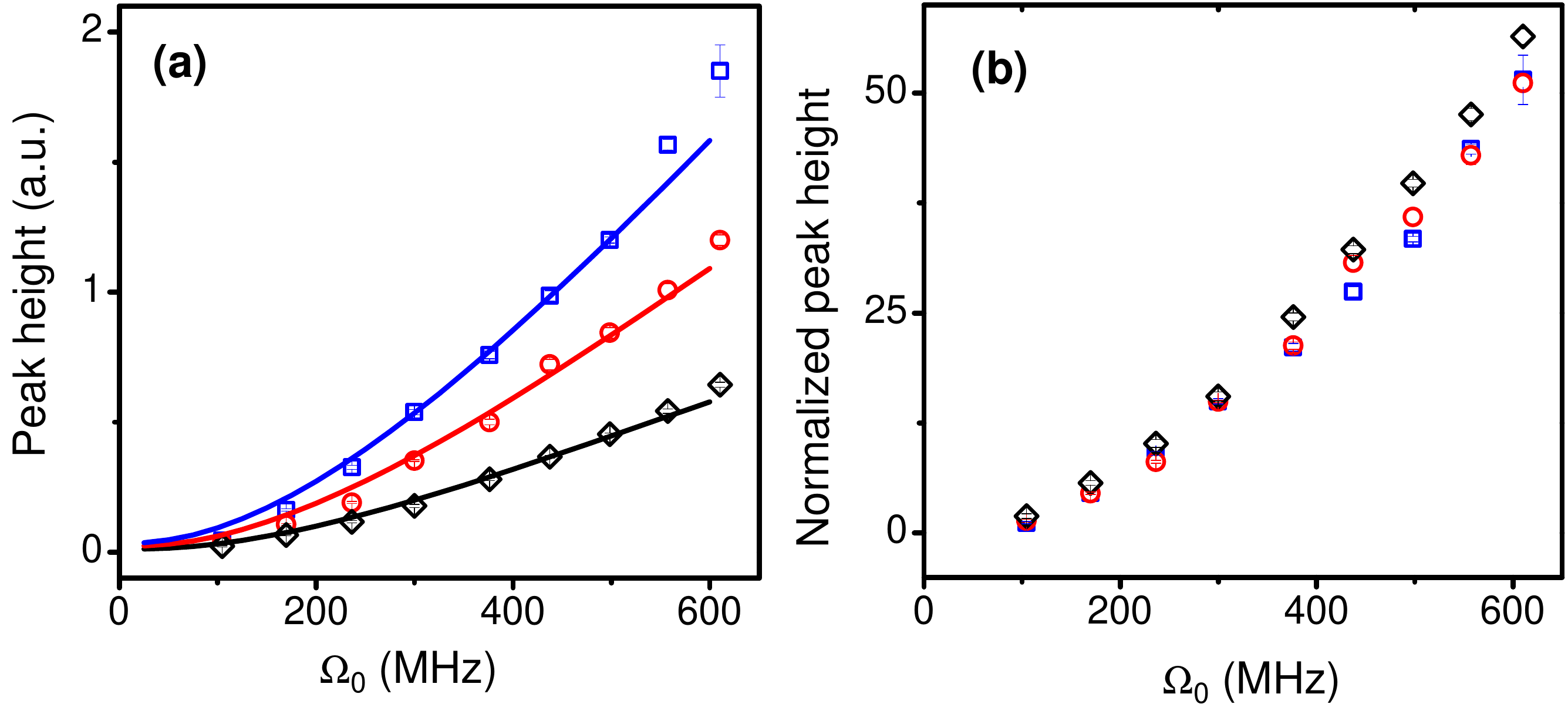,clip=,angle=0,width=8.5cm}
\caption[]{(a) Measured dispersion peak height as a function of peak Rabi frequency of probe ($\Omega_0$) while coupling to the Rydberg state ($n=30$s$_{1/2}$) with atomic vapor densities $2.5\times 10^{12}$ cm$^{-3}$ $(\diamond)$, $1.25\times 10^{13}$ cm$^{-3}$ $(\circ)$ and $3.0\times 10^{13}/$ cm$^{-3}$ $(\square)$. The peak Rabi frequency of the coupling beam was $13.8$ MHz. The solid lines are the fitting using the theoretical model and a multiplication factor is used as the only fitting parameter. (b) Dispersion peak height normalized to the peak height of a weak probe beam.}
\end{center}
\label{fig3}
\end{figure}

In a further study of variation of the dispersion peak height as a function of probe Rabi frequency, an RF attenuator at the output 
of the detectors was used to keep the amplitude of the beat constant irrespective of the probe laser power.
The variation of the dispersion peak height as a function of probe Rabi frequency for different vapor densities is shown in
figure 3(a). The dispersion peak height calculated by the theoretical model for the same laser parameters and vapor densities agrees 
well as shown in figure 3(a). When the dispersion peak height data is normalized to that of a weak probe beam, then all the
data corresponding to different densities fall on the same line as shown in figure 3(b). This observation suggests that the refractive 
index of the medium depends linearly on the vapor density and the Rydberg-Rydberg interaction has negligible effect.

\begin{figure}[t]
\begin{center}
\epsfig{file=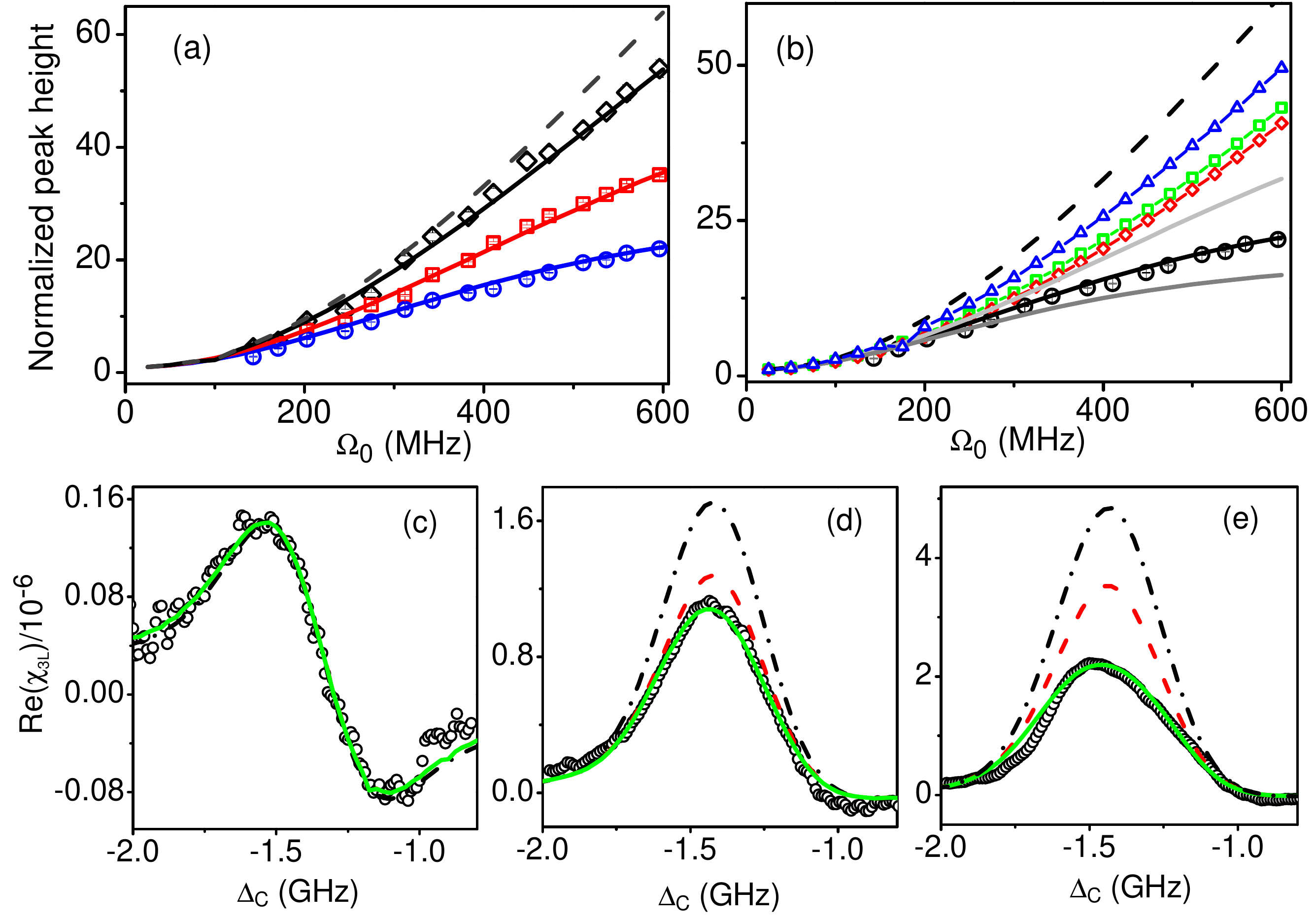,clip=,angle=0,width=8.5cm}
\caption[]{(a) Normalized dispersion peak height as a function of peak Rabi frequency of the probe beam ($\Omega_0$) coupling to the Rydberg state ($60$s$_{1/2}$) with atomic vapor densities $2.5\times 10^{12}$ cm$^{-3}$ $(\diamond)$, $1.25\times 10^{13}$ cm$^{-3}$ $(\square)$ and $3.0\times 10^{13}/$ cm$^{-3}$ $(\circ)$. The peak coupling Rabi frequency was $8.5$ MHz. The solid lines are the fitting using the theoretical model which includes interaction induced dephasing and Rydberg blockade. The dashed line is derived from the model without interaction. A multiplication factor and blockade radius are adjusted to compare with the experimental result. (b) The details of the dispersion peak height as function of peak probe Rabi frequency for atomic vapor density $3.0\times 10^{13}/$ cm$^{-3}$. The dashed line is derived from the model without interaction. The data points are generated using the model which includes interaction induced dephasing only, but without blockade and with $\Gamma_{rr}=100$ MHz $(\triangle)$, $\Gamma_{rr}=500$ MHz $(\square)$, and $\Gamma_{rr}=1000$ MHz $(\diamond)$. The solid lines are derived from the model which includes both interaction induced dephasing and blockade with $\Gamma_{rr}=500$ MHz and $r_b=1.65$ $\mu$m (light gray), $r_b=2.2$ $\mu$m (black), and $r_b=2.75$ $\mu$m (dark gray). The curve with $r_b=2.2$ $\mu$m matches well with the experimental data $(\circ)$. The dispersion spectra for the Rydberg state ($60$s$_{1/2}$) for the peak probe Rabi frequencies equal to $50$ MHz (c), $275$ MHz (d) and $510$ MHz (e). The green solid, red dashed, and black dotted dashed lines are the dispersion spectra generated using the model with both interaction induced dephasing and blockade, only with interaction induced dephasing but without blockade, and without interactions, respectively.}
\end{center}
\label{fig4}
\end{figure}  

To study the blockade interaction, the blue laser was tuned to interact with a Rydberg state with principal quantum number 
$n=60$ and the same experiment was performed. The variation of the dispersion peak height at different densities in the strong 
Rydberg-Rydberg interaction regime is shown in figure 4(a). In contrast to the observation presented in figure 3, the normalized 
dispersion peak height shows a non-linear dependence of density and there is a clear indication of suppression of the dispersion 
peak at higher densities. Since $Re\left(\chi_{3L}\right)\propto\rho_{rr}$, then suppression in dispersion peak height is due to the 
suppression of Rydberg population which is the signature of the Rydberg blockade interaction. The ions inside the vapor cell has 
negligible effect which was confirmed using Rydberg EIT~\cite{moha07} for $60$s$_{1/2}$ Rydberg state. The blackbody radiation 
induced ionization and transition rates are less than $10$ kHz~\cite{bete07} and has negligible effect.
To exclude the possibility of suppression in the dispersion peak due to interaction induced dephasing, we introduced a Rydberg
population dependent dephasing of the dipole matrix element $\rho_{rg}$ similar to the model discussed in reference~\cite{balu13}.
Rydberg population ($\rho_{rr}$) decays at a rate of $\Gamma_{rg}$ and $\rho_{rg}$ decays at a rate of $\frac{\Gamma_{rg}}{2}+\gamma_{rel}+
\rho_{rr}\Gamma_{rr}$. Introducing this term in the effective 2-level model, a cubic equation of $\rho_{rr}$ is obtained.
Looking at the coefficients of the cubic equation, one can deduce that out of the 3 solutions of $\rho_{rr}$, one is positive real
and the other two solutions are either negative real or complex conjugate to each other. The positive solution of $\rho_{rr}$ is 
evaluated by solving the cubic equation numerically and is replaced in equation (1) to calculate the dispersion of the probe beam. 
The dispersion peak height calculated using this model is shown in figure 4(b). We have observed that depending on $\Gamma_{rr}$, 
the dispersion peak height reduces in comparison with the non-interacting model, but monotonically increases as the probe Rabi 
frequency and doesn't display any feature of saturation. Also, by increasing $\Gamma_{rr}$ from $500$ MHz to $1$ GHz, a very small 
reduction of the dispersion peak is observed.  

To explain the saturation of the experimental data, we introduced blockade classically. Consider the number of atoms per blockade 
sphere to be $N_b$ nad blockade radius to be $r_b$. The probability of simultaneous multiple excitations of $n$ out of $N_b$ atoms 
to the Rydberg state is given by $P_n=\frac{N_b!}{n!(N_b-n)!}\rho_{rr}^n\left(1-\rho_{rr}\right)^{\left(N_b-n\right)}$. 
For these events, only one Rydberg excitation out of $n$ atoms is considered and the blockaded
Rydberg population is evaluated as $\rho_{rr}^{(b)}=\left(P_0+\sum_1^{N_b}\frac{P_n}{n}\right)\rho_{rr}$. The dispersion of the probe is then determined by 
replacing $\rho_{rr}^{(b)}$ in equation (1). The theoretical curves generated after introducing blockade into the model are shown in 
figure 4. If we take $r_b=2.2$ $\mu$m, then it fits well with the experimental data for all three densities. As shown in figure 4(b), 
if the blockade radius is changed by $25\%$, it clearly doesn't fit to the experimental data. 
The width of the dispersion spectra for different probe Rabi frequencies are shown in figure 4(c,d and e). 
For $\Omega_0=50$ MHz, the Rydberg population is small and in this non-interacting regime, the model with blockade and population 
dependent dephasing shows very little deviation from the non-interacting model. For $\Omega_0=275$ MHz, the model with population 
dependent dephasing reduces significantly from the non-interacting model but further including blockade doesn't show much deviation. 
For $\Omega_0=510$ MHz, there is a significant suppression of the peak due to blockade. However, the width of the dispersion spectrum 
doesn't change appreciably compared to the non-interacting model as shown in figure 4(e) which indicates a clear evidence of Rydberg 
blockade in our system. 
              
In conclusion, Rydberg blockade is demonstrated in thermal vapor and the blockade radius is measured to be $2.2$ $\mu$m. The van
der Waals coefficient ($C_6$) for the Rydberg state ($60$s$_{1/2}$) is $140$ Ghz/$\mu$m$^6$ ~\cite{prit10}. The typical width of the 2-photon 
resonance is about $500$ MHz which can be determined from figure 1(c). Hence, blockade radius in this system should be approximately
$2.5$ $\mu$m which differ from our experimental result by less than $15\%$. It is to be noted that we have introduced blockade 
classically and a full quantum mechanical model may give a better estimate for the blockade radius. Our result shows that the coherent
collective Rydberg excitation is possible in thermal vapor and opens up the possibility to build the quantum devices like single photon 
source and photonic phase gate based on Rydberg blockade non-linearity in thermal vapor.

\section{ACKNOWLEDGMENTS}
We acknowledge the fruitful discussions with Sabyasachi Barik and Surya N Sahoo regarding the heterodyne detection technique. We also thank 
Sushree S Sahoo for assisting in performing the experiment. This experiment was financially supported by the Department of Atomic
Energy, Govt. of India.
%\clearpage


\begin{thebibliography}{99}
\bibitem{rabi} The probe Rabi frequency is estimated using the relation with the intensity of probe as 
$I/I_{sat}=2\left|\Omega_p\right|^2/\Gamma^2$. For $^{85}$Rb, $I_{sat}=1.64$ mW and $\Gamma=2\pi\times6$ MHz. Coupling Rabi frequency 
is estimated by fitting the Rydberg EIT signal with a weak probe beam without focusing and then scaling it with the intensity of the beam.
\end{thebibliography}
\end{document}